\documentclass[aip,jap,twocolumn,10pt,longbibliography,showpacs,superscriptaddress,]{revtex4-1}
\usepackage{graphicx}
\usepackage{amsfonts}
\usepackage{amsmath}
\usepackage{natbib}

\begin{document}

\title{Robust operation of a GaAs tunable barrier electron pump}

\author{S.~P.~Giblin}
\affiliation{National Physical Laboratory, Hampton Road, Teddington, Middlesex TW11 0LW, United Kingdom}
\author{M.~-H.~Bae}
\author{N.~Kim}
\affiliation{Korea Research Institute of Standards and Science, Daejeon 34113, Republic of Korea}
\author{Ye-Hwan Ahn}
\affiliation{Korea Research Institute of Standards and Science, Daejeon 34113, Republic of Korea}
\affiliation{Department of Physics, Korea University, Seoul 02841, Republic of Korea}
\author{M.~Kataoka}
\affiliation{National Physical Laboratory, Hampton Road, Teddington, Middlesex TW11 0LW, United Kingdom}

\email[stephen.giblin@npl.co.uk]{Your e-mail address}

\date{\today}

\begin{abstract}
We demonstrate the robust operation of a gallium arsenide tunable-barrier single-electron pump operating with 1 part-per-million accuracy at a temperature of $1.3$~K and a pumping frequency of $500$~MHz. The accuracy of current quantisation is investigated as a function of multiple control parameters, and robust plateaus are seen as a function of three control gate voltages and RF drive power. The electron capture is found to be in the decay-cascade, rather than the thermally-broadened regime. The observation of robust plateaus at an elevated temperature which does not require expensive refrigeration is an important step towards validating tunable-barrier pumps as practical current standards.
\end{abstract}

\pacs{1234}

\maketitle

\section{Introduction}
The controlled transport of single electrons in mesoscopic devices has attracted much attention as a conceptually simple primary standard of electric current \cite{pekola2013single}. Very precise control of electrons has been achieved using chains of mescoscopic normal metal islands \cite{keller1996accuracy}, but limited to slow pumping rates $\lesssim 10$~MHz due to the fixed $RC$ time-constant of the junctions between the islands. At the present time, the most practically useful combination of accuracy and high electron pumping rate has been achieved using electrostatically gated semiconductor quantum dots (QDs) operated as non-adiabatic tunable-barrier pumps\cite{kaestner2015non} in the low-temperature decay cascade regime \cite{kashcheyevs2010universal}. Using state-of-the-art current measurement techniques \cite{giblin2012towards,drung2015ultrastable}, there have been several reports of pumped current accurate at the part-per-million (ppm) level or better, at pump repetition rates in the range $0.5$~GHz $\leq f \leq 1$~GHz, generating current $80$~pA $\leq I_{\text{P}}=ef$ $\leq 160$~pA \cite{giblin2012towards,bae2015precision,stein2015validation,yamahata2016gigahertz,stein2016accuracy,stein2016robustness}, where $e$ is the elementary charge. These studies were performed on a variety of device architectures: etch-defined\cite{giblin2012towards,stein2015validation,stein2016accuracy} and gate-defined \cite{bae2015precision} QDs in GaAs heterostructures, and silicon nano-wire MOSFETs \cite{yamahata2016gigahertz}. While very promising for the metrological application of electron pumps, most of these studies were performed on carefully tuned devices. The required robustness of the current against changes in the pump control parameters has only recently begun to be investigated with high precision \cite{stein2016robustnessCPEM,stein2016robustness}, and only in one type of etch-defined pump.

\begin{figure}
\includegraphics[width=8.5cm]{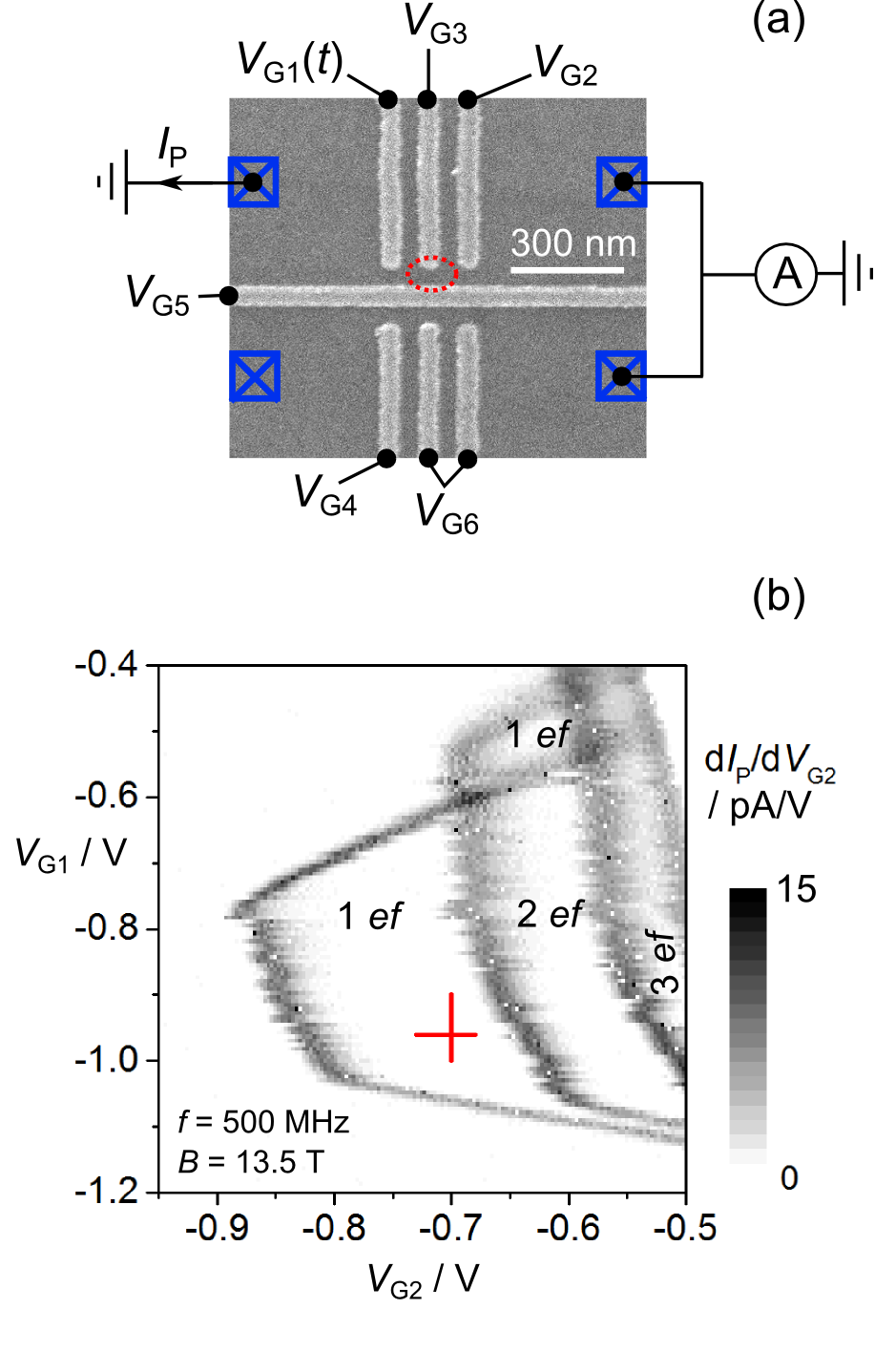}
\caption{\textsf{(a): Scanning electron microscope image of a device from the same fabrication batch as the one used in this study. Crossed boxes indicate ohmic contacts. Metallic gates show up as light grey. The connection of the $6$ control voltages to the gates is indicated, and when the gates are energized the QD forms in the approximate location indicated by the red dashed oval. An AC voltage added to gate $1$ pumps electrons from left to right, generating a current $I_{\text{P}}$ with conventional sign flowing from right to left. (b): Derivative map of pumped current at $f=500$~MHz, $B=13.5$~T and $P_{\text{RF}}=5.2$~dBm as a function of $V_{\text{G1}}$ and $V_{\text{G2}}$. The current on the quantised plateaus (white regions of the plot) is indicated in multiples of $ef$. The red lines indicate the ranges of the high-accuracy scans shown in Figs. 2a,b,e,f.}}
\label{fig:fig1}
\end{figure}

In this study, we broaden the study of robustness, and investigate the gate-defined tunable barrier pump \cite{seo2014improvement,bae2015precision}. Most significantly for the application of pumps as practical current standards, we perform our measurements at $\sim 1.3$~K, the temperature of pumped helium-4. This is in contrast to previous robustness studies \cite{giblin2010accurate,stein2016robustnessCPEM,stein2016robustness} which were carried out at dilution refrigerator temperatures. Using a rigorous statistical approach to evaluate the plateau extension and flatness, we find robust plateaus in all the tuning parameters we investigated, flat to within the $\sim 2\times 10^{-6}$ relative statistical uncertainty of each data point. Long measurements with the device in an optimally-tuned condition gave a current equal to $ef$ within a relative total uncertainty of $8.6\times 10^{-7}$. We also show that despite the elevated temperature, the pump was operating in the decay-cascade regime and not the thermally-broadened regime predicted \cite{kashcheyevs2014modeling} and observed \cite{yamahata2014accuracy} at higher temperatures. Furthermore, the device was affected by a significant amount of charge noise. The robust performance of the pump under these non-ideal conditions is encouraging evidence that the semiconductor electron pump can fulfill a role as a practical current standard. 

This paper is structured as follows: Section II describes the characterization and measurement technique. Section III presents the main experimental results in which we show that the pump current displays flat plateaus over a wide range of several tuning parameters. In section IV we analyze the statistical fluctuations of the current on the plateaus, and show that there is no indication of structure on the plateaus within the measurement uncertainty. Finally in section V we show that the pump is operating in the decay-cascade regime, and not in the thermal-equilibrium regime, even at the elevated temperature. 

\section{characterisation}
The pump used in this study (see SEM image in Fig. 1a) was realised in a 2-dimensional electron gas (2-DEG) in a GaAs-AlGaAs heterostructure with metallic surface gates. The sample was fabricated using techniques described previously \cite{seo2014improvement,bae2015precision}, and measured at a temperature of $\sim 1.3$~K. DC voltages $V_{\text{G1}}-V_{\text{G6}}$ defined a quantum dot in the region between the gates, and a sinusoidal AC voltage at $f=500$~MHz was added to gate $1$ using a room-temperature bias-T, to pump electrons from the source to the drain. The AC source had an output power $P_{\text{RF}}$, calibrated for a $50$~$\Omega$ load, and the total attenuation of the $50 $~$\Omega$ co-axial line between the source and the device was $\approx 4$~dB. A magnetic field $B=13.5$~T was applied perpendicular to the plane of the sample\cite{wright2008enhanced,kaestner2009single,fletcher2012stabilization}. The pump current $I_{\text{P}}$ was measured in two modes; normal-accuracy and high-accuracy. In normal-accuracy mode, used for rapid characterization, the current was amplified by a room-temperature transimpedance amplifier, with an uncertainty in the gain calibration of $\sim 2 \times 10^{-4}$. For high-accuracy measurements, $I_{\text{P}}$ was compared with a reference current derived from applying a voltage across a calibrated $1$~G$\Omega$ standard resistor. \cite{giblin2012towards,bae2015precision,yamahata2016gigahertz}. In this mode the amplifier measures the small difference between the pump and reference currents, and provided this difference is made less than $0.05 \%$ of $I_{\text{P}}$, by tuning the reference current, the calibration uncertainty of the amplifier contributes less than $1 \times 10^{-7}$ to the total relative uncertainty. We are chiefly interested in the deviation of $I_{\text{P}}$ from its expected quantised value $ef$, so we define the dimensionless normalised deviation,  $\Delta I_{\text{P}} \equiv (I_{\text{P}}-ef)/ef$.Likewise, all uncertainties in $\Delta I_{\text{P}}$ will be expressed as relative uncertainties in dimensionless units. The RF modulation of the entrance gate, and the reference current source are turned on and off synchronously with a cycle time of $40$~seconds to eliminate instrumental offsets. The on-off cycle is repeated $n_{\text{cyc}}$ times. To reject linear drift in the offset current, our data analysis routine calculates $\Delta I_{\text{P}}$ using the data from the 'off' part of the cycle and half of the data from the two adjacent 'on' parts, thus generating $n_{\text{cyc}}-1$ statistically independent values of $\Delta I_{\text{P}}$ with standard deviation $\sigma_{\text{I}}$ \footnote{see supplementary figure S1}. These values are then averaged to yield a mean $\Delta I_{\text{P}}$ with statistical uncertainty $U_{\text{ST}}=\sigma_{\text{I}}/ \sqrt{n_{\text{cyc}}-1}$ (all uncertainties reported in this paper are 1 sigma standard uncertainties). The relative systematic uncertainty in $\Delta I_{\text{P}}$ is dominated by the calibration uncertainty of the standard resistor, $U_{\text{1G}}=8 \times 10^{-7}$, with an additional small contribution due to the voltage measurement $U_{\text{V}}\lesssim 2 \times 10^{-7}$ so that the total uncertainty $U_{\text{T}}=\sqrt{U_{\text{ST}}^{2}+U_{\text{1G}}^{2}+U_{\text{V}}^{2}}$. 

Fig. 1b shows the derivative $dI_{\text{P}}/V_{\text{G2}}$ as a function of $V_{\text{G1}}$ and $V_{\text{G2}}$, obtained from a normal-accuracy measurement, following an iterative tuning procedure to find the optimum settings for the DC gate voltages: $(V_{\text{G1}},V_{\text{G2}},V_{\text{G3}},V_{\text{G4}},V_{\text{G5}},V_{\text{G6}}) = (-0.96, -0.7, 0.39, -0.78, 0.53, -1)$~V, and $P_{\text{RF}}=5.2$~dBm. During the tuning procedure, plots of $I_{\text{P}} (V_{\text{G1}},V_{\text{G2}})$ similar to Fig. 1b were obtained first while systematically stepping $V_{\text{G3}}$ and $V_{\text{G5}}$, with the aim of maximising the width of the $1ef$ plateau. At minimum, a $4\times4$ matrix of $(V_{\text{G3}},V_{\text{G5}})$ values were investigated. Having found the optimal values of $V_{\text{G3}}$ and $V_{\text{G5}}$, the procedure was repeated stepping $V_{\text{G4}}$ and $V_{\text{G6}}$. Note that the relatively large negative values of the voltages applied to the lower finger gates in Fig. 1a, combined with the positive voltage applied to the plunger gate $V_{\text{G3}}$, has the effect of shifting the QD position above the axis of symmetry defined by the trench gate $V_{\text{G5}}$. The approximate location of the QD is indicated by a dashed red circle in Fig. 1a\cite{seo2014improvement}.

The data of Fig. 1b was taken as a series of $V_{\text{G2}}$ scans at fixed $V_{\text{G1}}$, with $V_{\text{G1}}$ incremented between scans. This plot, known as the 'pump map', shows clearly the regions of zero derivative, where the current is invariant in the two control voltages \cite{blumenthal2007gigahertz,kaestner2008robust}. The mis-alignment of regions of maximum derivative in successive scans visible in this data also shows that the device operation is affected by a random telegraph signal (RTS) well known to affect this type of 2-DEG structure \cite{cobden1991noise,liefrink1994low} and already observed in another sample \cite{bae2015precision} with a similar design to the one in this study. Despite the noise, a broad region can be identified on the one-electron plateau where the derivative is zero within the resolution of the data. In the next section, we use high-accuracy measurements to investigate the robustness of current quantization on the one-electron plateau.

\begin{figure*}
\includegraphics[width=17cm]{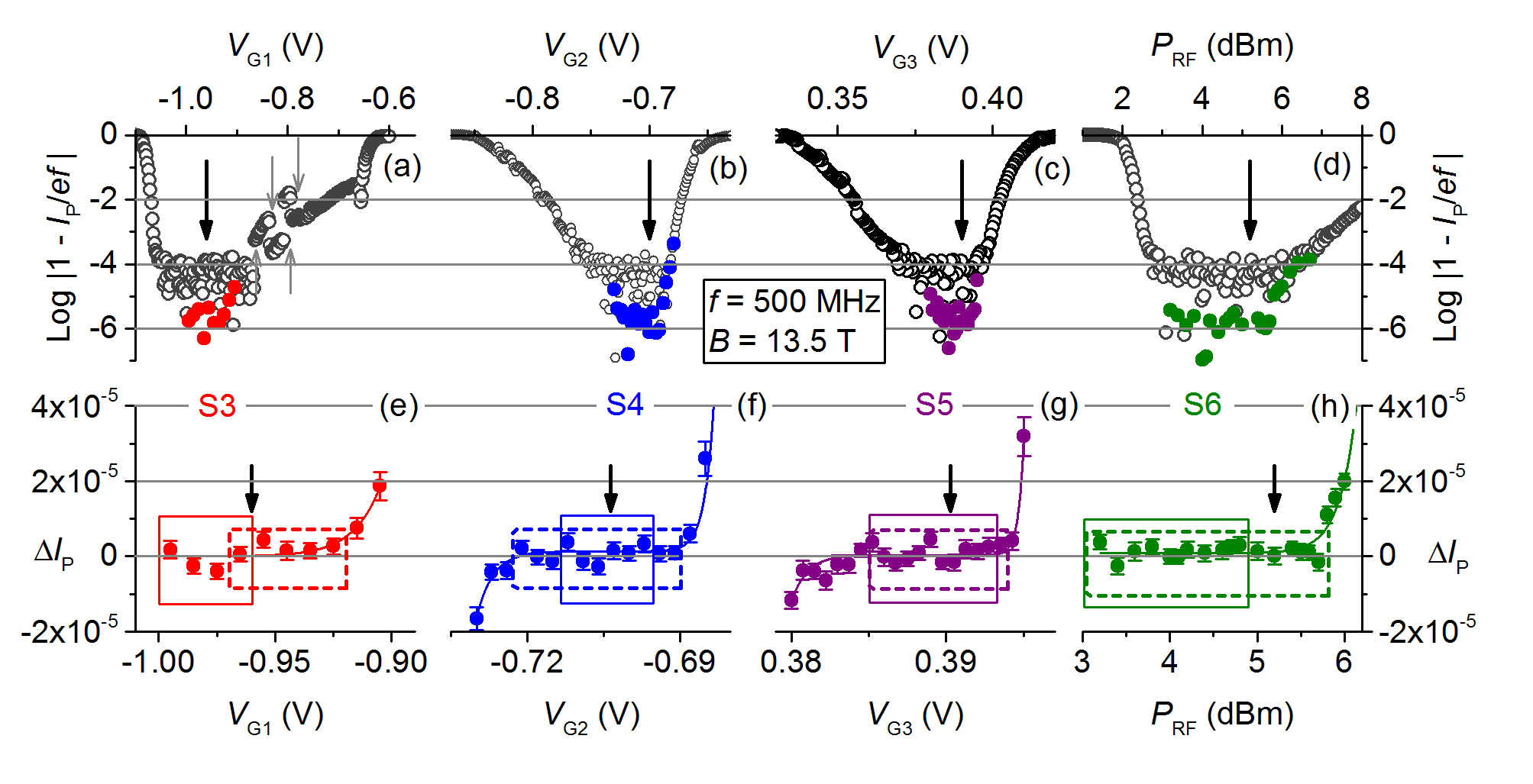}
\caption{\textsf{Pumped current at $f=500$~MHz and $B=13.5$~T as a function of four control variables (left to right) $V_{\text{G1}}$, $V_{\text{G2}}$, $V_{\text{G3}}$, and $P_{\text{RF}}$. The upper panels (a-d) show normal-accuracy (open circles) and high-accuracy (filled circles) data on a log scale. Lower panels (e-h) show the same high-accuracy data as the upper panels, on a linear scale with error bars indicating the statistical uncertainty. Solid lines in the lower plots are fits to equation 1. Boxes show the range of data for which a plateau can be defined: solid boxes by reference to the exponential fit lines of equation 1, and dashed boxes by reference to linear fits (linear fit lines are not shown). Thick vertical arrows on all the plots indicate the optimum values of tuning parameters $(V_{\text{G1}},V_{\text{G2}},V_{\text{G3}}) = (-0.96, -0.7, 0.39)$~V, and $P_{\text{RF}}=5.2$~dBm. The thin vertical arrows in panel (a) indicate RTS events.}}
\label{fig:fig2}
\end{figure*}

\section{high-accuracy plateau measurements}
We made a total of $6$ high-accuracy measurement scans as a function of the control parameters $V_{\text{G1}}$, $V_{\text{G2}}$, $V_{\text{G3}}$ and $P_{\text{RF}}$, denoted S1-S6, as well as normal-accuracy measurements over a wider range of each scanned parameter. We also made a further $4$ measurements with the pump tuning parameters fixed to the optimal values and $n_{\text{cyc}}=750$, $1400$, $900$ and $983$, denoted F1-F4. The six scans and four fixed-parameter measurements were made over a period of $14$ days. In Fig. 2 we present data from four of the scans, with each set of high-accuracy data plotted (filled circles) on logarithmic (Figs. 2(a-d)) and linear (Figs. 2(e-h)) axes, with normal-accuracy data (open circles) also shown on the logarithmic plots. Each high-accuracy data point in the data of Fig. 2 is averaged from $70$ on-off cycles. The error bars indicate the statistical uncertainty $U_{\text{ST}} \sim 2 \times 10^{-6}$, which for these relatively short averaging times is the largest component of the total uncertainty; $U_{\text{T}} \sim U_{\text{ST}}$. The normal-accuracy data has sufficient accuracy and signal-to-noise ratio to resolve relative deviations of $\Delta I_{\text{P}}$ from $ef$ as small as $10^{-4}$, and the logarithmic plot is a useful way to visualize the data during the iterative gate tuning procedure. In each scan plotted in Fig. 2, the fixed parameters were set to the optimum values noted in section II. Two additional scans were performed, S1 and S2 (not shown in Fig. 2), with one fixed parameter slightly offset from the optimum: S1 was a $V_{\text{G2}}$ scan, with $V_{\text{G1}} = -0.975$~V, and S2 was a $V_{\text{G1}}$ scan with $V_{\text{G2}}=-0.695$~V.

The effect of RTS noise can be seen in the normal-accuracy data, particularly for scan S3, where individual RTS switching events are indicated by gray arrows in Fig. 2a. Nevertheless, for each scan, the high-accuracy data exhibits a plateau where $\Delta I_{\text{P}}$ appears invariant in the control parameter within the uncertainty of the individual data points. Scans S3 and S4 can immediately be compared with similar data measured using an etch-defined pump \cite{stein2016robustness}, and we note that the plateaus in our gate defined pump are approximately twice as wide in both entrance gate ($V_{\text{G1}}$) and exit gate ($V_{\text{G2}}$) as those in the etch-defined pump. This may reflect a higher charging energy of the gate-defined pump, but it could also be an artifact of different lever arms (gate voltage to QD energy conversion factors) resulting from the very different geometries of the two types of device. Comparing scans S4 and S5 (Fig. 2f,g) the effect of the different lever arms of $V_{\text{G2}}$ and $V_{\text{G3}}$ on the QD level is clear: both of these gates control the depth of the QD, so $I_{\text{P}}$ has a similar functional dependence on either gate, but because $V_{\text{G3}}$ is coupled much more strongly to the QD than $V_{\text{G2}}$, the plateau occupies a smaller range of gate voltage.

To evaluate the plateau extension more quantitatively, two methods were used. Firstly (the 'exponential fit method'), we fitted the high-accuracy data $I_{\text{P}}(x)$ to a sum of two exponential functions \cite{kashcheyevs2014modeling}
\begin{equation}
\frac{I_{\text{FIT}}}{ef} = 1+ \delta_{\text{I}} -e^{-\alpha_{1} (x-x_{1})} + e^{\alpha_{2} (x-x_{2})}
\end{equation}
where $\alpha_{\text{1}}$, $\alpha_{\text{2}}$, $x_{\text{1}}$, $x_{\text{2}}$, $\delta_{\text{I}}$ are fitting parameters. The parameter $\delta_{\text{I}}$ is the best-fit offset of the plateau from $ef$. We include it because we do not assume \textit{a priori} that the plateau is exactly quantised. For runs S1-S6, we found $0.23\times 10^{-6} \leq \delta_{\text{I}} \leq 1.33\times 10^{-6}$. For runs S3 and S6, only the second exponential term was used for the fit because the data had no clear deviation from the plateau on the low-x axis side. We defined the plateau width as the range of the control parameter for which $|(I_{\text{FIT}}/ef)-1-\delta_{\text{I}}| \leq \delta_{\text{fit}}$, with $\delta_{\text{fit}}=10^{-7}$. This choice of $\delta_{\text{fit}}$ reflects the lower limit to the statistical uncertainty achievable for realistic measurement times of order $1$ day. Other studies\cite{stein2015validation,stein2016robustness} used the same method to define the plateau, but without including the offset $\delta_{\text{I}}$, and with $\delta_{\text{fit}}=10^{-8}$. The fits are shown in the lower panels of Fig. 2 as solid lines\footnote{See supplementary figure S1 for an example illustration of the fit line on an expanded y-axis}, and the resulting selections of data points (number of points $=N_{\text{exp}}$) are enclosed by a solid box. The standard deviation of the $N_{\text{exp}}$ data points in each scan is denoted $\sigma (\Delta I_{\text{P}})$, and the statistical uncertainty of $\Delta I_{\text{P}}$ averaged over these points on the plateau is $U_{\text{ST,plat}}=\sigma (\Delta I_{\text{P}})/ \sqrt{N_{\text{exp}}}$. The scatter of the data points inside the boxes appears to be consistent with their individual uncertainties, but we will address this point more quantitatively in section IV.

\begin{table*}
\caption{Fit and slope parameters}
\centering
\setlength{\tabcolsep}{8pt}
\begin{tabular}{c c c c c c c c c}
\hline\hline
Scan number & Scanned variable & $n_{\text{cyc}}$ & $N_{\text{exp}}$ & $N_{\text{lin}}$ & plateau width & Slope $S$ & $U_{\text{SLOPE}}$ & Flatness \\[0.5ex]
 &  &  &  &  &  & $\times 10^{-6}$ & $\times 10^{-6}$ & $\times 10^{-6}$ \\[0.5ex]
\hline
S1 & $V_{\text{G2}}$ & 25 & 15 & 21 & 40 mV & 74 /V & 81 /V & 3.24 \\
S2 & $V_{\text{G1}}$ & 70 & 6 & 11 & 100 mV & 19 /V & 23.7 /V & 2.37 \\
S3 & $V_{\text{G1}}$ & 70 & 4 & 5 & 40 mV & 18 /V & 53 /V & 2.12 \\
S4 & $V_{\text{G2}}$ & 70 & 6 & 11 & 30 mV & 24 /V & 59 /V & 1.77 \\
S5 & $V_{\text{G3}}$ & 70 & 11 & 12 & 8.25 mV & 233 /V & 261 /V & 2.15 \\
S6 & $P_{\text{RF}}$ & 70 & 11 & 17 & 2.5 dBm & -0.22 /dBm & 0.51 /dBm & 1.27 \\[1ex]
\hline
\end{tabular}
\end{table*}

\begin{figure}[!]
\includegraphics[width=8.5cm]{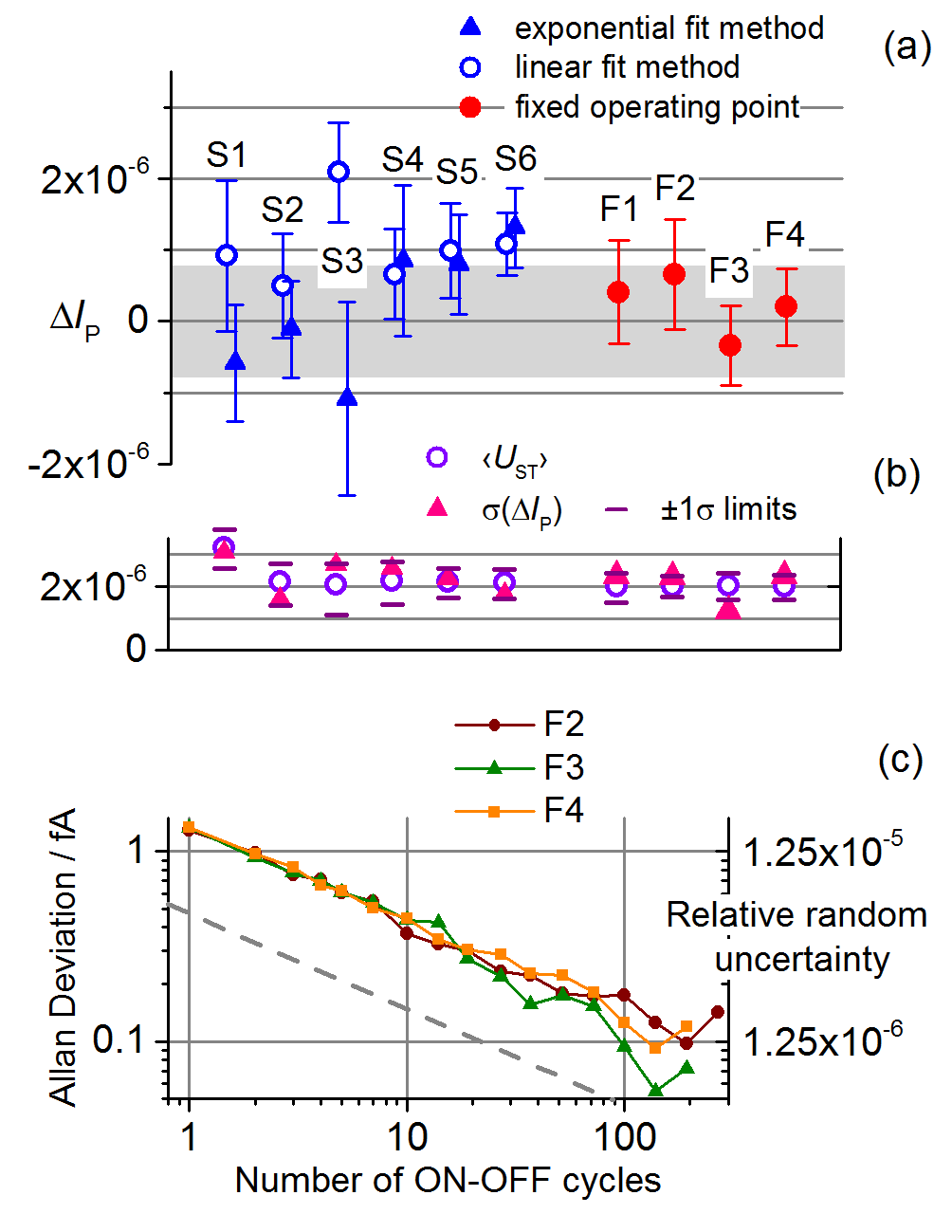}
\caption{\textsf{(a): Average pumped current at $f=500$~MHz and $B=13.5$~T obtained from each experimental run, with the run number indicated above the data points. The error bars are the sum of the un-correlated uncertainty components $\sqrt{U_{\text{ST,plat}}^2 + U_{\text{V}}^2}$ for scans S1-S6, and $\sqrt{U_{\text{ST}}^2 + U_{\text{V}}^2}$ for fixed-points runs F1-F4. The uncertainty in the resistor calibration, $U_{\text{1G}}=8\times 10^{-7}$, is shown as a shaded grey box. For scans S1-S6, the plotted value is an average over a range of a control parameter, with the range selected using the linear fit method (open circles) and exponential fit method (filled triangles). (b): Statistical properties of the data in plot (a). Data points corresponding to the same run are aligned vertically in (a) and (b). Open circles show the mean $U_{\text{ST}}$ for the data points on-plateau, as selected by the exponential fit method for runs S1-S6. For runs F1-F4, they show the mean $U_{\text{ST}}$ for the data set analyzed in blocks of $70$ on-off cycles. Filled triangles show the standard deviation of the data points on-plateau, and for runs F1-F4 the standard deviation of the data analyzed in blocks of $70$ cycles. Horizontal bars show the $68 \%$ coverage upper and lower bounds for $N_{\text{exp}}$ measurements of $\Delta I_{\text{P}}$ to have a given standard deviation, assuming that $\Delta I_{\text{P}}$ is normally distributed with standard deviation $\langle U_{\text{ST}} \rangle$. (c): Allan deviation of pumped current as a function of the number of on-off cycles, calculated from $3$ of the runs at fixed operating point. The Allan deviation for run F1 (not shown) exhibited similar behavior. The gray dashed line shows the expected $1/\sqrt{t} $ dependence for frequency-independent Johnson-Nyquist noise in the reference resistor.}}
\label{fig:fig3}
\end{figure}

Secondly, a purely empirical criterion was used, based on linear fits to sections of the high-accuracy data (the 'linear fit method'). This method does not make any assumptions about the functional form of the data. For each scan, we found the largest number $N_{\text{lin}}$ of consecutive data points for which $|S|<U_{\text{SLOPE}}$, where $S$ is slope of a linear fit to the $N_{\text{lin}}$ points, and $U_{\text{SLOPE}}$ is the uncertainty in the slope\footnote{An estimate of the uncertainty in the slope is given by the square root of the second diagonal element in the $2 \times 2$ covariance matrix output by the fitting algorithm. See standard texts on regression analysis, for example Draper and Smith, \textit{Applied Regression Analysis} (second edition), Wiley (1981), pp 82-85.}. The resulting data ranges are enclosed by dashed boxes in the lower panels of Fig. 2, and the relevant parameters are shown in table 1. As with the exponential fit method, the statistical uncertainty of the averaged points is given by $U_{\text{ST,plat}}=\sigma (\Delta I_{\text{P}})/ \sqrt{N_{\text{lin}}}$. The linear fit method allows us to assign a numerical value to the plateau flatness given by $U_{\text{SLOPE}}$ multiplied by the plateau width. The flatness is comparable to the uncertainty of the data points from which it is derived, because the scatter of the data points determines the uncertainty in the linear regression. The flatness therefore is roughly between $1 \times 10^{-6}$ and $2 \times 10^{-6}$ for all the scans irrespective of the plateau width in the scanned units. For example, scans S4 and S5 have plateau widths in gate voltage units differing by roughly a factor $3$ due to the different lever arms of $V_{\text{G2}}$ and $V_{\text{G3}}$ as noted above, but the flatness for both the plateaus is $\sim 2 \times 10^{-6}$. To evaluate the flatness with $10^{-7}$ uncertainty using the linear fit method would require long averaging times, but we note that this is the only unambiguous method of proving that a plateau is flat. The exponential fit method, on the other hand, allows the plateau extension to be estimated based on a much shorter measurement, under the strong assumption that the fitting function (in this case, an exponential) captures all of the physics relevant to the pump accuracy at the target level of uncertainty. 

For all the scans, $N_{\text{exp}}<N_{\text{lin}}$, which is to be expected since we chose $\delta_{\text{fit}} \ll U_{\text{ST}}$; the exponential fit method estimates the plateau extension to be smaller than the linear fit method, because the latter is only constrained by $U_{\text{ST}} \sim 2 \times 10^{-6}$. For scan S3, scatter of some of the data points strongly constrained the range of points which satisfied the linear fit criterion. As can be seen from table I, a similar scan, S2, exhibited a plateau in $V_{\text{G1}}$ more than twice as wide in gate voltage. The question of whether the scatter in run S3 is excessively large is addressed in section IV. Regarding the $P_{\text{RF}}$ scan S6, there are some indications in Fig. 2h that an exponential function does not adequately describe the increase of the current for $P_{\text{RF}} > 5.7$~dBm, and we speculate that rectification \cite{giblin2013rectification} or heating may play a role in the breakdown of quantised pumping at large gate drive amplitudes. 

The current averaged over the plateaus, with the plateaus defined using both the exponential (closed triangles) and linear (open circles) fit methods, is plotted in Fig. 3a for runs S1-S6. Error bars show the un-correlated uncertainty $\sqrt{U_{\text{ST,plat}}^2 + U_{\text{V}}^2}$. The current measured in runs F1-F4 with the pump at fixed operating point is also plotted on the same graph (closed circles), with error bars indicating $\sqrt{U_{\text{ST}}^2 + U_{\text{V}}^2}$. The un-correlated uncertainty does not include $U_{\text{1G}}$, which is shown as a grey box centred on $\Delta I_{\text{P}}=0$. The resistor was calibrated before and after the measurement campaign and its value was assumed constant during the campaign based on its long-term drift rate of $\sim 0.01 (\mu \Omega / \Omega)$/day \cite{giblin2012towards}. In contrast, the voltage measurement was calibrated before and after each run. The un-correlated uncertainty thus allows the different measurements of $\Delta I_{\text{P}}$ to be compared with each other without the additional uncertainty associated with linking to the SI unit system. For example, the two fixed-point runs with the lowest uncertainty, F3 and F4, are consistent within their combined uncertainty of $7.8 \times 10^{-7}$. If the plateau is defined using the exponential fit method, the average current is consistent with $ef$ within the uncertainties, and furthermore there are no major inconsistencies between the data points when only the un-correlated uncertainty is considered. Averaging all the data from the four fixed-point runs (a total of $4033$ cycles lasting $47$ hours) reduced $U_{\text{ST}} $ such that $U_{\text{T}} \sim U_{\text{1G}}$ and yielded a best estimate of the pump current: $\Delta I_{\text{P}}=0.28 \pm 0.86 \times 10^{-6}$. This is marginally more accurate than the previous best electron pump measurement using the current measurement system at NPL\cite{yamahata2016gigahertz}, although it falls short of the record low uncertainty of $1.6 \times 10^{-7}$ recently reported \cite{stein2016robustness} using a measurement system based on a new type of ultra-stable current preamplifier known as an 'ULCA' \cite{drung2015ultrastable}. Future efforts will aim to reduce $U_{\text{1G}}$ to around $2\times 10^{-7}$ as well as implementing an ULCA-based measurement system at NPL. It is interesting to note that the accumulated precision measurements and associated theoretical fit lines \cite{giblin2012towards,bae2015precision,stein2015validation,yamahata2016gigahertz,stein2016robustness}, suggest that a tunable-barrier electron pump operated at an optimal working point is accurate at the $1 \times 10^{-7}$ level. With this premise, we could hypothetically consider the pump as a primary current standard, and the data of runs F1-F4 as constituting a calibration of the reference resistor with total uncertainty $\sim 3\times 10^{-7}$, almost a factor $3$ lower than the $U_{\text{1G}}$ presently achievable at NPL. However, we believe such a step would be premature, and that the robustness of these pumps requires further extensive investigation before a consensus can be reached on the required set of conditions for operation at a given accuracy level. 

\section{statistical evaluation of plateau current}

The data points on the plateaus in Fig. 2 show some scatter, and we now evaluate whether this scatter is consistent with statistical scatter about a stationary mean or whether it is a sign of structure on the plateau, or possibly drift in the pump current or the measurement system. We note that recent developments in the metrology of small currents \cite{drung2015ultrastable,drung2015validation} have focused attention on the stability of high-value thick-film standard resistors, principally those of $100$~M$\Omega$ value. The $1$~G$\Omega$ standard resistor used in the reference current source is also a thick-film design, and may suffer from short-term instability at the sub-ppm level. However, the uncertainties in the data of Figs. 2 and 3 are too large for this to have a significant effect on the scatter of the data points. We focus on the more conservative (narrower) plateaus defined using the exponential fit method. For each scan, the mean of the $N_{\text{exp}}$ values of the statistical uncertainty $U_{\text{ST}}$, denoted $\langle U_{\text{ST}} \rangle$, is plotted as the open points in Fig 3b. We also plot as solid points the standard deviation $\sigma (\Delta I_{\text{P}})$ of the $N_{\text{exp}}$ values of $\Delta I_{\text{P}}$. If the current on the plateau was drifting on the time-scale of the scan, or if the plateau was not flat, we expect $\sigma (\Delta I_{\text{P}}) > \langle U_{\text{ST}} \rangle$. To assign a statistical significance to the ratio $\sigma (\Delta I_{\text{P}}) / \langle U_{\text{ST}} \rangle$, we used a numerical simulation to assign a $68 \%$ confidence interval to the distribution of $\sigma (\Delta I_{\text{P}})$ expected for $N_{\text{exp}}$ normally-distributed measurements with standard deviation $\langle U_{\text{ST}} \rangle$\footnote{See supplementary information for more detail on the statistical analysis}. This is plotted as upper and lower horizontal bars in Fig. 3b. The fixed-parameter runs F1-F4 were evaluated in the same way as the scans, by dividing the data into blocks of $70$ cycles and analyzing each block separately. Over the whole data set, there is no statistically significant deviation of the ratio $\sigma (\Delta I_{\text{P}}) / \langle U_{\text{ST}} \rangle$ from $1$. One particular run, S3, appeared to have anomalously large scatter, visible in Fig. 2(e) and already discussed in section III. This scatter is apparent in Fig. 3(b), in the relatively large ratio of $\sigma (\Delta I_{\text{P}}) / \langle U_{\text{ST}} \rangle$. However, $\sigma (\Delta I_{\text{P}})$ is still just within the $68 \%$ confidence interval, clarifying that the data at different $V_{\text{G1}}$ values cannot be distinguished from data drawn from the same distribution. Overall, we conclude from this analysis that the scatter of the data points on the plateaus is consistent with statistical fluctuations about a stationary mean. 

This conclusion is supported by the Allan deviation of the current measured from runs F1-F4, all of which exhibited similar behavior. The Allan deviation plots for runs F2-F4 are shown in Fig. 3c. They show no significant deviation from the expected $\sqrt{t}$ behavior for frequency-independent noise out to the longest averaging times probed by the Allan deviation analysis \cite{allan1987should}, roughly one quarter of the total measurement time, or $\sim 3$~hours. For comparison, the dashed line shows the expected Allan deviation of frequency independent Johnson-Nyquist noise in the $1$~G$\Omega$ resistor, $(4.2$~fA/$\sqrt{\text{Hz}}) / (\sqrt{2 \tau})$, where $\tau = 40$~s is the time for one on-off cycle. The Allan deviation of the pump current is increased above this theoretical level due to three inefficiencies in the duty cycle which reduce the effective averaging time: The on-off cycle means the pump current is only measured for half the time, auto zero in the readout voltmeters halves the measurement time again, and rejection of data points at the start of each half-cycle, to eliminate transient effects, further reduces the duty cycle. The latter two of these effects need to be optimized in future experiments to yield a lower overall statistical uncertainty\cite{stein2016robustness}.

\section{pumping regime and noise broadening}

The relatively high temperature of these measurements compared to previous high-precision studies motivated us to consider the mechanism of charge capture by the pump. At low temperatures, this occurs by a cascade of one-way tunneling events whereby electrons tunnel back to the source electrode as the QD is progressively isolated from the source \cite{fujiwara2008nanoampere,kashcheyevs2010universal}. The experimental signature of the decay cascade is a characteristic double-exponential shape to the pump current as a function of the QD depth-tuning parameter. This tuning parameter can be the 'exit gate' voltage in simple two-gate pumps \cite{giblin2010accurate,giblin2012towards,stein2015validation}, or a global top gate voltage \cite{fujiwara2008nanoampere,yamahata2016gigahertz}, and in this work its role can be fulfilled by either $V_{\text{G2}}$ or $V_{\text{G3}}$. At higher temperatures, experimental \cite{yamahata2014accuracy} and theoretical \cite{fricke2013counting,kashcheyevs2014modeling} work has indicated a cross-over to a thermal regime, in which back-tunneling is accompanied by forward tunneling into the QD from the source. This results in a symmetric shape to the current as a function of QD depth tuning parameter, reflecting the Fermi distribution of electrons in the leads. The cross-over to the thermal regime has been predicted to occur for $10 \times k_{\text{B}}T \gtrsim \Delta_{\text{ptb}}$ \cite{kashcheyevs2014modeling}. Here, $\Delta_{\text{ptb}}$ is defined as the change in energy of the QD level when the entrance barrier transmission changes by a factor of Euler's number $\sim 2.718...$ and it thus quantifies the device-specific cross coupling between the modulated entrance barrier, and the QD energy level \cite{fricke2013counting}. We crudely estimate $\Delta_{\text{ptb}} \sim 1$~meV$=8.9\times k_{\text{B}}T$ for our device, based on the slope of representative conductance pinch-off data and typical lever arm factors between a gate voltage and QD energy level. From this estimate we expect the device to be between the two regimes, and we next examine experimental data to clarify the capture mechanism.

\begin{figure}
\includegraphics[width=8.5cm]{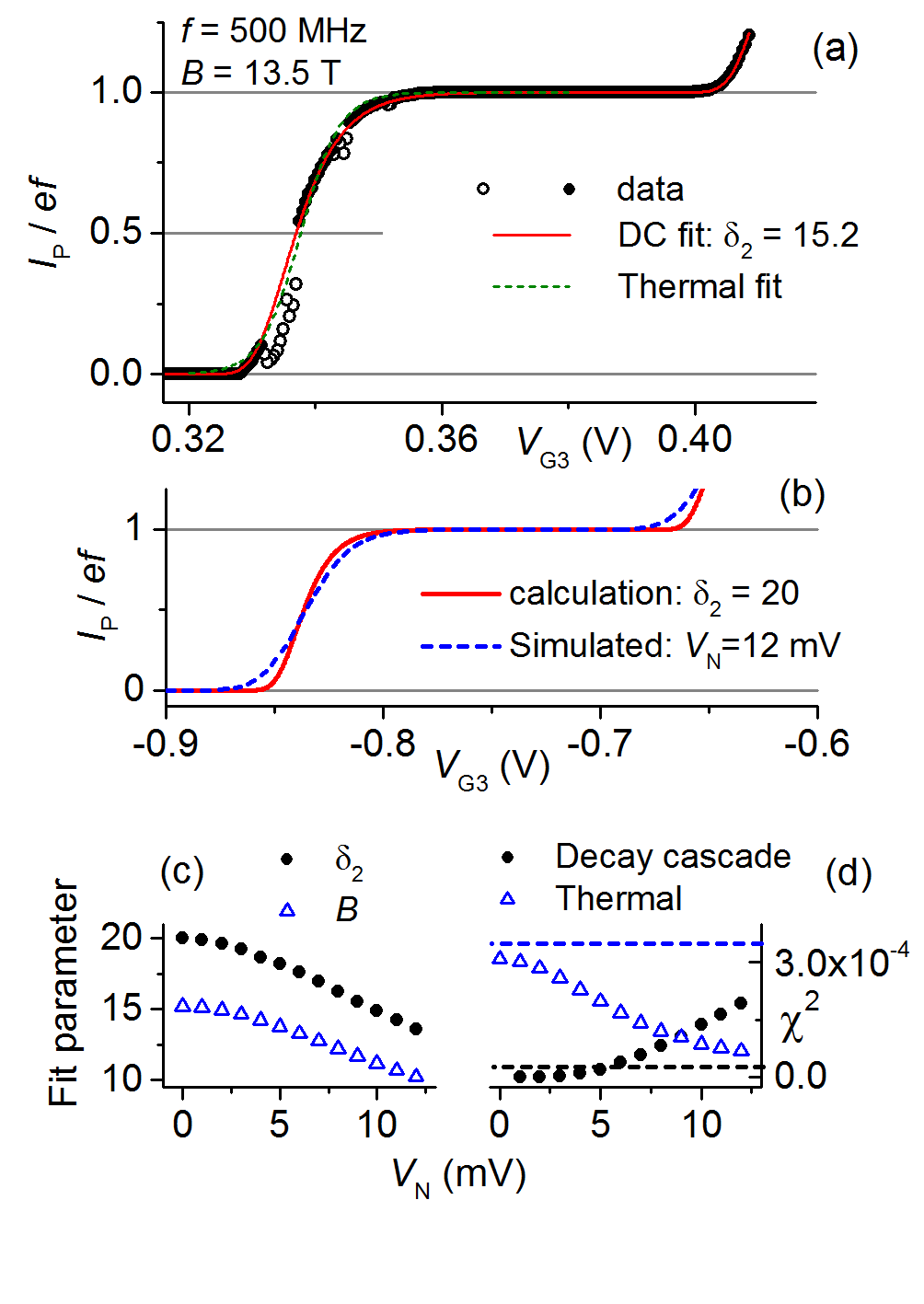}
\caption{\textsf{(a): Open and filled points: normalized pump current as a function of $V_{\text{G3}}$ measured using the normal-accuracy method. $f=500$~MHz, $B=13.5$~T and $P_{\text{RF}}=5.2$~dBm. Solid line: fit of equation (2) to the filled points. Dashed line: fit of equation (3) to the filled points. Open points denote data excluded from the fits. (b): Plot of equation (2) (solid line) and after simulated broadening by Gaussian noise with standard deviation $12$~mV (dashed line). (c) and (d) show fit parameters (c) and Chi-Square (d) as a function of noise amplitude for fits of equations (2) (filled circles) and (3) (open triangles) to simulated noise-broadened data similar to that shown as the dashed line in plot (b).}}
\label{fig:fig4}
\end{figure}

In Fig. 4a, we plot the normalized pump current as a function of $V_{\text{G3}}$, which functions as a QD depth-tuning gate. A RTS is visible in the transition between the plateaus, where the pump current is a sensitive probe of changes in the electrostatic potential. On the plateau, the current is insensitive to the state of the RTS. For the data of Fig. 4a, in the transition region between $I_{\text{P}}=0$ and $I_{\text{P}}=ef$, the charge state causing the RTS noise appears to be in one state for the majority of the data points (filled points), and the points affected by a switch to the other state (open points) were excluded from fitting. The data is fitted to the decay cascade model \cite{kashcheyevs2010universal}: 
\begin{equation} \frac{I_{\text{P}}}{ef}=\displaystyle\sum\limits_{m=1}^2 \exp(-\exp(-aV_{\text{G3}}+\Delta_{m}))\end{equation} 
over the full range (solid line), with reduced $\chi^{\text{2}}=2.6\times 10^{-5}$, yielding the fit parameter $\delta_{\text{2}}\equiv (\Delta_{\text{2}}-\Delta_{\text{1}})=15.2$, and a thermal equilibrium (Fermi function) model \cite{fricke2013counting,yamahata2014accuracy}: 
\begin{equation}  \frac{I_{\text{P}}}{ef}=\frac{1}{1+e^{(A-V_{\text{G3}})/B}}\end{equation}
in the range $0.32 \leq V_{\text{G3}} \leq 0.38$ (dotted line) with reduced $\chi^{\text{2}}=3.5\times 10^{-4}$. Close inspection of the fit lines shows that the decay-cascade model gives a better fit, and the thermal equilibrium model fails to reproduce the asymmetric plateau shape, with a sharp riser from $I_{\text{P}}=0$ and a more gradual approach to $I_{\text{P}}=ef$. The reduced $\chi^{\text{2}}$ for the decay-cascade fit is more than a factor $10$ smaller than the thermal equilibrium fit, suggesting that the pump is operating in the decay cascade regime. A similar conclusion was drawn by fitting equations (2) and (3) to a $I_{\text{P}}(V_{\text{G2}})$ scan \footnote{see supplementary figure S3(a)} which was obtained with a faster sweep rate to the data of Fig 4(a). This data was not so much affected by RTS switching events, at the expense of a much smaller number of data points ($\sim 15$) in the region between $I_{\text{P}}=0$ and $I_{\text{P}}=ef$. For this data, equation (2) yielded $\delta_{\text{2}}=15.6$ with $\chi^{\text{2}}=2.6\times 10^{-4}$ and equation (3) yielded a fit with $\chi^{\text{2}}=6.0\times 10^{-4}$. 

The high RF power levels used in this experiment, corresponding to on-chip peak-to-peak gate voltages of order $1$~V, raise the possibility that the electron temperature in the leads is elevated from the refrigerator bath temperature, for example by RF currents from the entrance gate flowing to ground through stray capacitances and parts of the leads. We did not estimate the electron temperature in the leads, but some insight can be gained by cooling the device to $300$~mK. If RF-induced heating was the dominant mechanism determining the electron temperature at a bath temperature of $1.3$~K, we would not expect further reduction of the bath temperature to have any effect on the device characteristics. In fact, we observe a considerable sharpening of the plateau when the device is cooled to $300$~mK; fits of $I_{\text{P}}(V_{\text{G2}})$ to equation (2) yield $\delta_{\text{2}}\sim 20$, compared to $\delta_{\text{2}}\sim 15$ at $1.3$~K\footnote{see supplementary figures S3(a) and (b)}. We can conclude that RF-induced heating is not a dominant mechanism determining the device characteristics at a bath temperature of $1.3$~K, although it may play a role at $300$~mK. 

We also rule out the possibility that the data of Fig. 4a is broadened by noise leading to erroneous conclusions from the fits. We calculated numerically the effect of Gaussian fluctuations in $V_{\text{G3}}$, with standard deviation $V_{\text{N}}$, on the ideal decay-cascade behavior described by equation (2). Fig. 4b shows eq. (2) with $\delta_{\text{2}}=20$ (solid line), and after broadening with $V_{\text{N}}=12$~mV (dotted line). The broadened characteristic is more symmetric and resembles a thermal distribution. Fitting the noise-broadened characteristic to the decay-cascade formula results in a decreasing $\delta_{\text{2}}$ parameter as $V_{\text{N}}$ is increased (Fig. 4c, solid symbols), but also a progressive reduction in the quality of the fit, reflected in an increase in $\chi^{\text{2}}$ (Fig. 4d, solid symbols). Fitting to the thermal function, eq. (3), the reverse is true: the thermal fit becomes a more accurate description of the simulated data for larger noise amplitudes (Fig. 4d, open symbols). Comparing the actual $\chi^{\text{2}}$ values obtained from fitting the data of Fig. 4(a) (horizontal dashed lines in Fig. 4(d)) with those calculated from noise broadening, we conclude that the experimentally measured data is not consistent with more than a few mV of noise broadening, and the pump is indeed operating in the decay cascade regime at our experimental temperature.

\section{conclusions}
In conclusion, pumping in a GaAs tunable-barrier electron pump is robust against changes in the gate control parameters, and the RF drive amplitude, at the part-per-million level at a temperature of $1.3$~K. The presence of two-level flucutators did not affect the accuracy of the pump current. Compared to previous studies, this relaxes the experimental conditions required to observe quantised pumping at the part-per-million accuracy level, which is a promising step towards adoption of quantised charge pumps as current standards. 

\begin{acknowledgments}
This research was supported by the UK department for Business, Energy and Industrial Strategy, the Joint Research Project 'Quantum Ampere' (JRP SIB07) within the European Metrology Research Programme (EMRP)and the EMPIR Joint Research Project 'e-SI-Amp' (15SIB08). The EMRP is jointly funded by the EMRP participating countries within EURAMET and the European Union. The European Metrology Programme for Innovation and Research (EMPIR) is co-financed by the Participating States and from the European Union's Horizon 2020 research and innovation programme. The study was partially supported by the Korea Research Institute for Standards and Science (16011245) and the national Research  Foundation  of  Korea  (NRF-2016R1A5A1008184 and NRF-2016K1A3A7A03951913).
\end{acknowledgments}

\bibliography{SPGrefs_KRISSPump}

\section{supplementary information}

The purpose of this supplementary information is to provide more detail on the analysis process for the high-accuracy measurements (Figure S1), and the calculation of statistical quantities used in the main text (Figure S2). We also show a comparison between exit gate characteristics at two different temperatures (Figure S3), illustrating the sharpening of the plateau when the pump is cooled from $1.3$~K to $300$~mK.

\renewcommand{\thefigure}{S\arabic{figure}}
\setcounter{figure}{0} 
\begin{figure*}
\includegraphics[width=12cm]{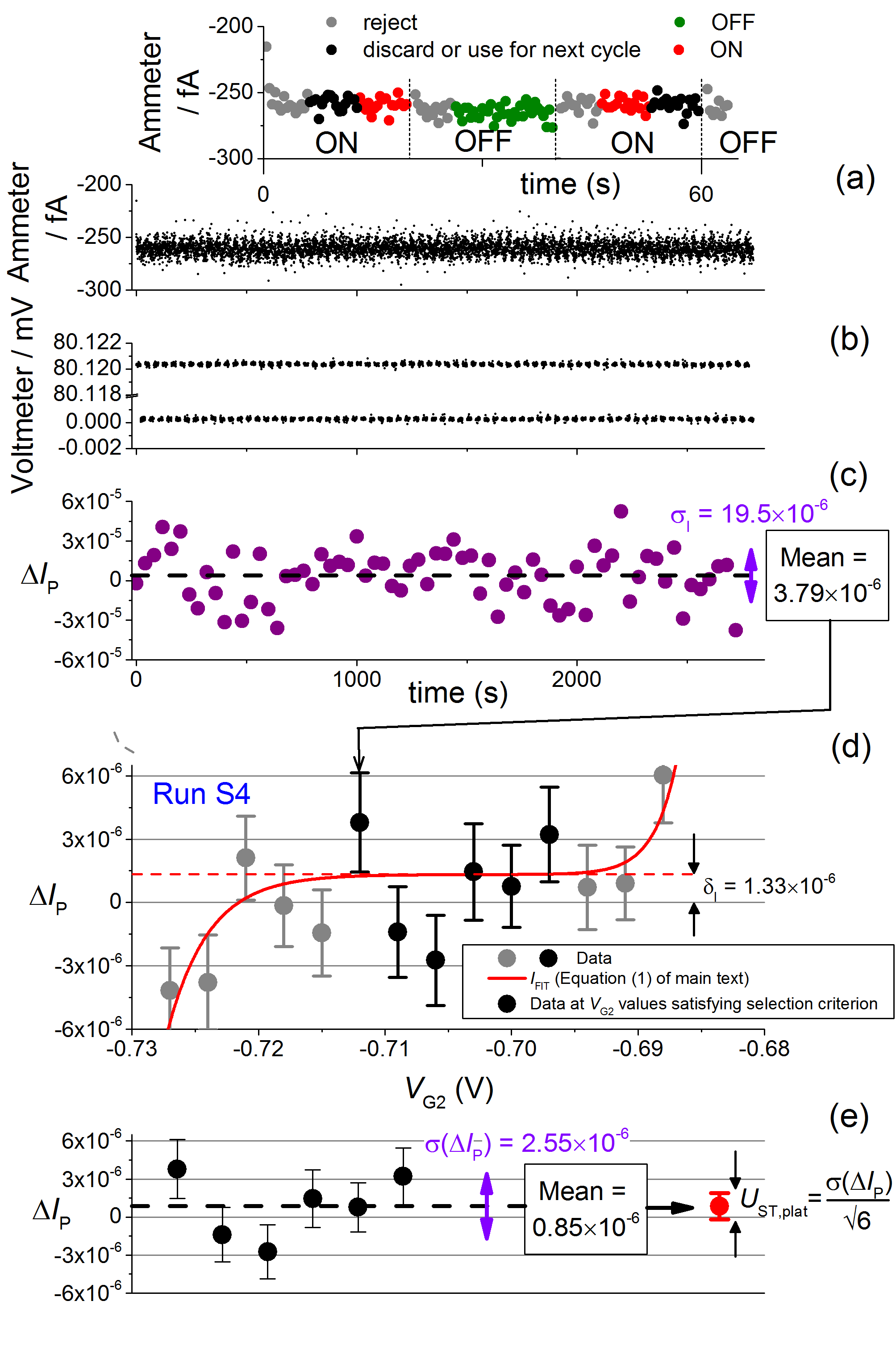}
\caption{\textsf{(a,b): Raw ammeter (a) and voltmeter (b) data. The data shows $n_{\text{cyc}}=70$ on-off cycles from run S4 at $V_{\text{G2}}=-0.712$~V. The inset to (a) shows the first $\sim 60$~s of data with the points color-coded to illustrate the data analysis protocol. The pump and reference current source are turned on and off synchronously every $40$~s, indicated by vertical dashed lines. (c): Values of $\Delta I_{\text{P}}$ obtained from processing the data raw data from panels (a) and (b). The mean of the $(n_{\text{cyc}}-1)=69$ points is indicated by a horizontal dashed line, and the standard deviation $\sigma_{\text{I}}$ is indicated by the vertical double-arrow. Panels (a)-(c) share the same time axis. (d): Reproduction of Figure 2(f) of the main text, showing scan S4 on an expanded y-axis. The data point indicated by the arrow is averaged from the data points of panel (c). As in Figure 2 of the main text, the error bars are the statistical uncertainty $U_{\text{ST}}=\sigma_{\text{I}}/ \sqrt{n_{\text{cyc}}-1}$ and the solid red line shows a fit to equation (1) of the main text. The fitting parameter $\delta_{\text{I}}$, which is the offset of the fit line from $\Delta I_{\text{P}}=0$, is indicated on the plot by a horizontal red dashed line. (e): (Black filled circles) Selection of $N_{\text{exp}}=6$ data points from panel (d) at $V_{\text{G2}}$ values for which $|(I_{\text{FIT}}/ef)-1-\delta_{\text{I}}| \leq 1 \times 10^{-7}$. The mean of these 6 data points is indicated by a horizontal dashed line, and the standard deviation by a vertical double arrow. The red data point to the right shows the mean with the error bar indicating statistical uncertainty $U_{\text{ST,plat}} = \sigma (\Delta I_{\text{P}})/ \sqrt{N_{\text{exp}}}$. This is the same data point plotted as a solid triangle labeled 'S4' in Fig. 3(a) of the main text, although in the main text the error bar is marginally larger because it includes $U_{\text{V}}$.}}
\label{fig:fig1}
\end{figure*}

Figure S1 illustrates the process of analysing raw data. As explained in the main text, the high-accuracy measurement system compares the unknown pump current $I_{\text{P}}$ with a reference current generated by applying a voltage across a calibrated $1$~G$\Omega$ resistor. The raw data are readings from two instruments: an ammeter which measures the difference between the pump and reference currents (Fig. S1(a)) and a voltmeter which measures the voltage across the $1$~G$\Omega$ resistor (Fig. S1(b)). The currents are switched on and off synchronously with a cycle time of $40$ seconds to remove instrumental offsets. Note that due to careful tuning of the reference current source, the (on-off) ammeter difference signal $\sim 10$~fA is only just visible in the raw data. In this study, each half-cycle consisted of $50$ readings from each instrument, triggered synchronously, with the instruments set to integrate for $10$ power line cycles (at a nominal power line frequency of $50$~Hz) and including an auto zero measurement with each reading. Thus, each instrument reading takes $0.4$~s, each half-cycle takes $20$~s and the $70$ cycles shown in the figure take $2800$~s. 

The inset to Fig. 1(a) shows a small portion of the ammeter data from the main panel; the first one and a half on-off cycles. The points are color-coded to illustrate the data analysis procedure. The first $16$ data points following each current switch (grey) are rejected to eliminate transient effects. The remaining $34$ data points from the \textit{off} half-cycle (green) are averaged to yield $I_{\text{OFF}}$. The $34$ points from each \textit{on} half-cycle are divided into two equal portions, and the two blocks of $17$ points adjacent to the \textit{off} cycle are averaged to yield $I_{\text{ON}}$. The ammeter difference signal extracted from the illustrated data is $\Delta I(1) = I_{\text{ON}}-I_{\text{OFF}}$. The first block from the first \textit{on} half-cycle (first section of black points) is discarded altogether, and the second block from the second \textit{on} half-cycle (second section of black points) is analysed with the second \textit{off} half cycle, and the first block from the third \textit{on} half-cycle to yield $\Delta I(2)$, and so on up to $\Delta I(n_{\text{cyc}}-1)$. The voltmeter data is analyzed in a similar way to yield $\Delta V(j)$, with $1 \leq j \leq (n-1)$, and the pump current is given by $I_{\text{P}}(j) = \Delta V(j)/R + \Delta I(j)$ \cite{giblin2012towards}. Breaking the data set up in this manner makes the measurement of $I_{\text{P}}$ insensitive to linear drift in the offset of the measured signals, at the expense of discarding data from one on-off cycle. The data analysis thus yields $n_{\text{cyc}}-1$ values of $\Delta I_{\text{P}}$ from a raw data set of $n_{\text{cyc}}$ cycles (Fig. S1(c)). The mean and standard deviation $\sigma_{\text{I}}$ of the data of Fig. S1(c) yield one data point in Fig. S1(d) indicated by an arrow, with $\Delta I_{\text{P}}=3.79\times 10^{-6}\pm U_{\text{ST}} =(3.79 \pm 2.33)\times 10^{-6}$. Here, the statistical uncertainty $U_{\text{ST}}$ is given by the standard error on the mean $\sigma_{\text{I}}/ \sqrt{n_{\text{cyc}}-1}$. For this data point the relative total uncertainty $U_{\text{T}}=\sqrt{U_{\text{ST}}^{2}+U_{\text{1G}}^{2}+U_{\text{V}}^{2}}=2.47\times 10^{-6} \sim U_{\text{ST}}$.

Referring to the scatter of the data points in Fig. S1(d), we define two more statistical terms, with reference to Fig. S1(e). Here, we have re-plotted the $N_{\text{exp}} = 6$ data points in Fig. S1(d) which are determined to be on the $ef$ plateau by using the exponential fit method (equation (1) of the main text). The mean of the data points is indicated by a horizontal dashed line. Each of the data points has a statistical uncertainty, and we calculate the mean of these statistical uncertainties, denoted $\langle U_{\text{ST}} \rangle = 2.18\times 10^{-6}$. We also calculate the standard deviation of the 6 data points, denoted $\sigma (\Delta I_{\text{P}})$, from which we derive the standard error of the mean $=\sigma (\Delta I_{\text{P}})/ \sqrt{N_{\text{exp}}} = U_{\text{ST,plat}}$. To recap, the symbol $U_{\text{ST}}$ denotes the statistical uncertainty for a measurement of $\Delta I_{\text{P}}$ at fixed pump operating point, while $U_{\text{ST,plat}}$ denotes the statistical uncertainty of an average of several measurements of $\Delta I_{\text{P}}$ at different operating points along a plateau.

If the plateau is truly flat, the scatter of the data points on the plateau given by $\sigma (\Delta I_{\text{P}})$ should be on average the same as the uncertainty $U_{\text{ST}}$ of a single data point. In other words, data points measured at different points along the plateau are sampling the same stationary mean value with a standard deviation given by the same underlying noise process. The dominant source of noise in our experiment comes from the measurement system: Johnson noise in the $1$~G$\Omega$ reference resistor, with additional small contributions from the current pre-amplifier and cryogenic wiring. This is reflected in the almost constant (within $\sim 10 \%$) values of $U_{\text{ST}}$ for $n_{\text{cyc}}=70$ visible in Fig. 3(b) (open circles) of the main text. We can therefore state that on a true plateau, data points with  $n_{\text{cyc}}=70$ should be drawn from a normal (Gaussian) parent distribution with standard deviation $\langle U_{\text{ST}} \rangle$. Since we measure a limited number $N_{\text{exp}}$ of data points on the plateau, we can compare the standard deviation $\sigma (\Delta I_{\text{P}})$ of these points with the expected distribution of the standard deviation, if we randomly selected $N_{\text{exp}}$ points from the parent distribution. To do this, we numerically generated a parent distribution with a large number of data points, and randomly selected $N_{\text{exp}}$ points from it. The random selection was repeated $10^{6}$ times, and in Fig. S2 (a), we plot the histogram of the $\sigma (\Delta I_{\text{P}})$ values for $N_{\text{exp}}=6$ and $\langle U_{\text{ST}} \rangle=2.18\times 10^{-6}$ (the parameters corresponding to run S4, Fig. 2(f) of the main text). The value of $\sigma (\Delta I_{\text{P}})$ measured for run S4 is shown as a dashed vertical white line super-imposed on the histogram. In the inset to Fig. S2(a) we plot a histogram of $(N-1)  [\sigma (\Delta I_{\text{P}})]^{2} / \langle U_{\text{ST}} \rangle ^{2}$ (black bars) for the same random data set as the main panel, and for comparison, the $\chi ^{2}$ distribution with $5$ degrees of freedom (solid red line). This illustrates a standard text-book result, namely that the variance of $N$ randomly selected points is distributed according to the $\chi ^{2}$ distribution with $N-1$ degrees of freedom. Fig. S2(b) shows the cumulative sum of the histogram in (a), with vertical dashed lines showing the $1\sigma$ (68 \% coverage) upper and lower limits to $\sigma (I_{\text{P}})$. These are plotted as horizontal bars for run S4 in Fig. 3(b) of the main text. The process illustrated in Fig. S2 was repeated with the parameters $[\langle U_{\text{ST}} \rangle, N_{\text{exp}}]$for the remaining $5$ scans to derive the horizontal bars in Fig. 3(b) of the main text. 

\begin{figure*}
\includegraphics[width=12cm]{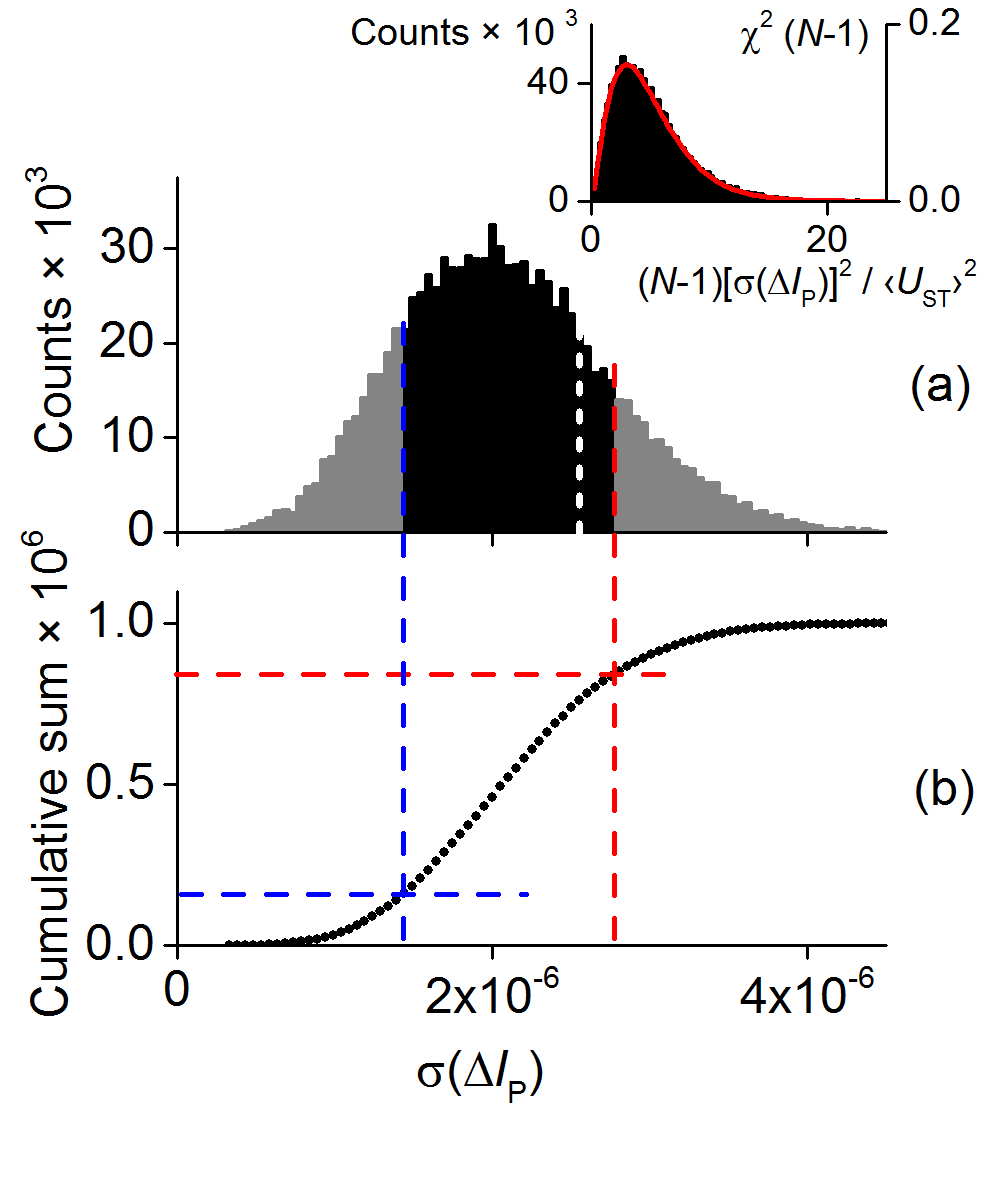}
\caption{\textsf{(a): Histogram of the standard deviation $\sigma (\Delta I_{\text{P}})$ of $N_{\text{exp}}=6$ data points randomly selected from a numerically generated normal distribution with standard deviation $\langle U_{\text{ST}} \rangle =2.18 \times 10^{-6}$. The parameters $[\langle U_{\text{ST}} \rangle, N_{\text{exp}}]$ correspond to the data points within the solid box for run S4 (Fig. 2f of the main text), and the standard deviation $\sigma (\Delta I_{\text{P}})$ for run S4 is shown as a white vertical dashed line. The random selection was repeated $10^{6}$ times. (b): cumulative sum of the histogram in (a). Horizontal dashed lines demarcate the conventional $1-\sigma$ upper and lower boundaries, and the intersection of these lines with the data gives the upper and lower $1-\sigma$ boundaries for a sample of $6$ points having a given standard deviation. These boundaries are indicated by vertical dashed lines which are extended up to plot (a), where the histogram bars are color-coded black (inside the $1-\sigma$ boundary) or gray (outside the $1-\sigma$ boundary). The inset to (a) (black bars, left y-axis) shows $(N-1)  [\sigma (\Delta I_{\text{P}})]^{2} / \langle U_{\text{ST}} \rangle ^{2}$, with $N=6$, for the same simulation data as the main panel. The solid red line (right y-axis) shows the $\chi ^{2}$ distribution with $(N-1) = 5$ degrees of freedom.}}
\label{fig:fig2}
\end{figure*}

\begin{figure*}
\includegraphics[width=17cm]{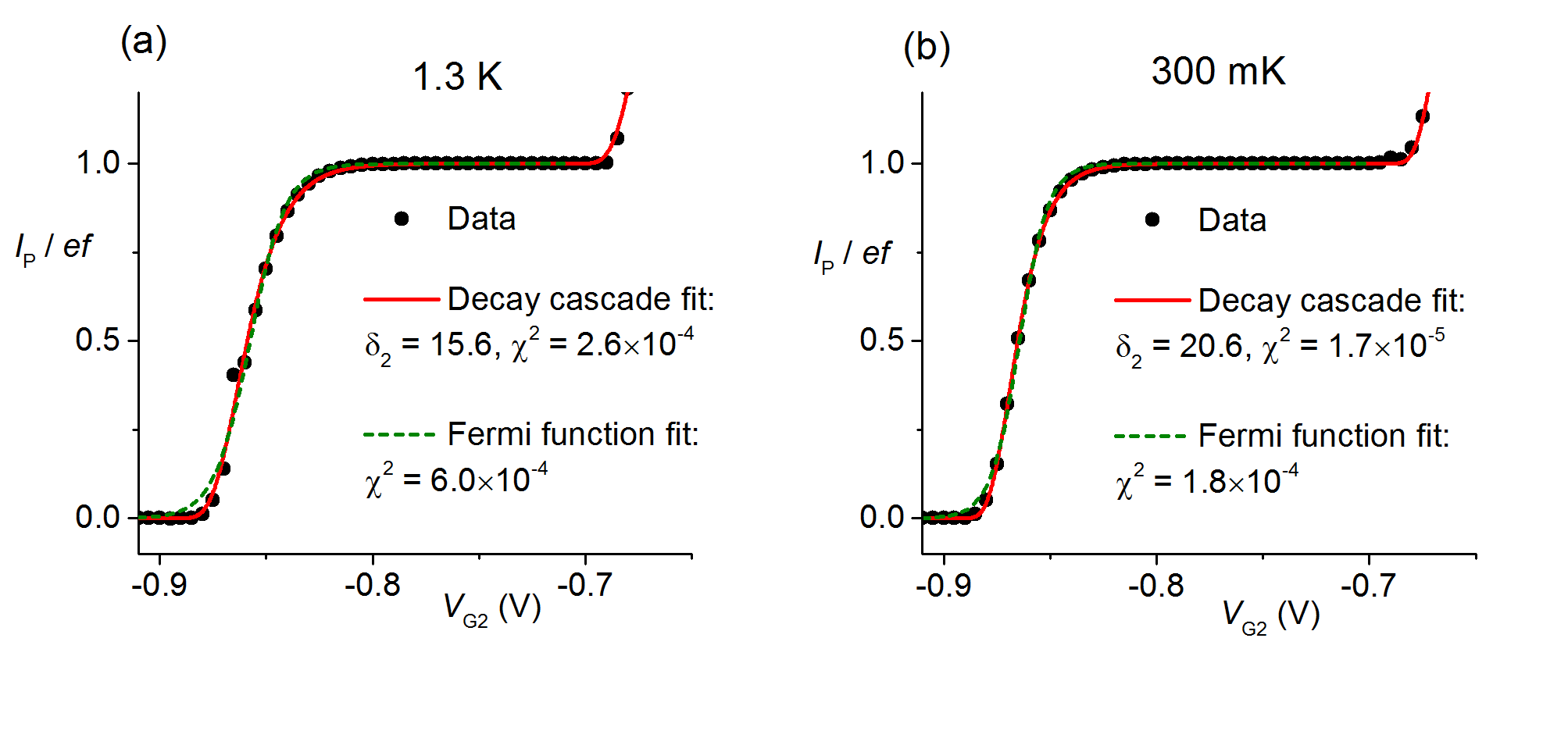}
\caption{\textsf{Normalised pump current as a function of $V_{\text{G2}}$ at temperatures of $1.3$~K in panel (a), and $300$~mK in panel (b). The gate voltages are tuned to the optimal working point used for the data of the main text, apart from $V_{\text{G1}}=-0.94$~V. Each data set has been fitted to the decay cascade model (equation (2) of the main text) and a Fermi function (equation (3) of the main text), with the independent variable $V_{\text{G3}}$ in the equations replaced by $V_{\text{G2}}$. $\chi ^{2}$ values for the fits are indicated on the plot. As for the $\chi ^{2}$ values reported in the main text, $\chi ^{2}$ is the sum of the square of the fit residuals divided by the number degrees of freedom.}}
\label{fig:fig3}
\end{figure*}

\end{document}